# Radiative Emission Enhancement Using Nano-antennas Made of Hyperbolic Metamaterial Resonators

Caner Guclu[1], Ting Shan Luk[2,3], George T. Wang[2], Filippo Capolino[1]

[1]Electrical Engineering and Computer Science, University of California, Irvine, Irvine , CA, 92697
[2]Sandia National Labs PO Box 5800, Albuquerque, NM, 87185
[3]Center for Integrated Nanotechnologies, Sandia National Laboratory, Albuquerque, NM 87185, USA

*Abstract*— A hyperbolic metamaterial (HM) resonator is analyzed as a nano-antenna for enhancing the radiative emission of quantum emitters in its vicinity. It has been shown that the spontaneous emission rate by an emitter near a hyperbolic metamaterial substrate is enhanced dramatically due to very large density of states. However, enhanced coupling to the free-space, which is central to applications such as solid-state lighting, has not been investigated significantly. Here, we numerically demonstrate approximately 100 times enhancement of the free-space radiative emission at 660 nm wavelength by utilizing a cylindrical HM resonator with a radius of 54 nm and a height of 80 nm on top of an opaque silver-cladded substrate. We also show how the free-space radiation enhancement factor depends on the dipole orientation and the location of the emitter near the subwavelength resonator. Furthermore, we calculate that an array of HM resonators with subwavelength spacings can maintain most of the enhancement effect of a single resonator.

*Index Terms*— hyperbolic metamaterial, nano-antennas resonator, plasmonics, spontaneous emission.

CONTROL of the spontaneous emission process is of fundamental and practical interest for enhanced light-matter interaction, quantum information processing, solid-state lighting, and biological sciences. Hyperbolic metamaterials (HMs) have been proposed for enhancing spontaneous emission rate owing to indefinite photonic density of states. HMs are characterized by an indefinite permittivity tensor (for example, $\underline{\boldsymbol{\varepsilon}}_r = \varepsilon_t \left( \hat{\mathbf{x}}\hat{\mathbf{x}} + \hat{\mathbf{y}}\hat{\mathbf{y}} \right) + \varepsilon_z \hat{\mathbf{z}}\hat{\mathbf{z}}$, where $\varepsilon_t \varepsilon_z$ is negative) and are in general made of metal-dielectric multilayers [1-3]; dielectric-semiconductor multilayers [4]; graphene-dielectric multilayers [5-7]; or by an array of metallic nanopillars [8,9]. HMs possess hyperbolic iso-frequency wavevector dispersion properties and can host ideally an infinite spectrum of propagating plane waves, which in turn leads to a dramatic increase in the photonic local density of states. The hyperbolic dispersion characteristic in a multilayer HM does not rely on building blocks with narrow-band resonances, therefore this property is maintained over a wide frequency band. Theoretical analysis of enhanced spontaneous emission due to a broadband Purcell effect in HMs is reported in [10,11] and the spontaneous emission lifetime reduction of dyes deposited on HM substrates have been experimentally observed[12-14]. Directional single photon emission has also been reported[15]. However, in those studies, the emitted power is mostly directed into the HM substrate where it is eventually dissipated as heat[1]. This leads to a very low external quantum efficiency even though there is a significant increase in the decay rate. In many enhanced emission applications, such as solid-state lighting, it is important to enhance the free-space radiative emission, i.e., the power escaping out of the system and not just the emitted one by the source. While enhanced free-space radiative emission has been shown in a HM grating structure[16], three-dimensional HM resonator properties have not been explored. Here we show an approximately 100-fold enhancement in free-space radiation emission using three-dimensional nanoscale HM subwavelength resonators.

It has been shown that HM resonators can possess high optical quality factor even in extreme subwavelength dimensions and that the radiated power by an excited HM resonator can be larger than the dissipated power[17,18]. However, free-space radiative emission enhancement due to HM resonators was not studied in [17,18]. Due to the anomalous resonance wavelength scaling with HM resonator size, one can adjust the resonator size without changing the resonance wavelength in contrast to standard dielectric or plasmonic resonators. These properties are unique to HM resonators and allow one to optimize the radiation efficiency and radiative emission enhancement when a sub-wavelength HM resonator is used as a nano-antenna. Here we explicitly show the strong enhancement of free-space *radiative* emission and quantum efficiency when a quantum emitter is placed close to a HM resonator.

In the following we first discuss the optical properties of an isolated cylindrical HM resonator (comprising silver-silica multilayers) located on top of an opaque silver-cladded substrate. Then we report radiation enhancement for varying dipole position and orientation. Finally, we elaborate on the radiation enhancement produced by a finite two-dimensional array of HM resonators with subwavelength spacings. The reported enhancement opens up a possibility for enhancing solid state lighting performance.

We consider a quantum dot (QD) emitter located very close to the HM resonator. We model the QD emitter at various locations relative to the resonator as a dipole with polarization oriented along the principal axes. A schematic view of the configuration is shown in Fig. 1.

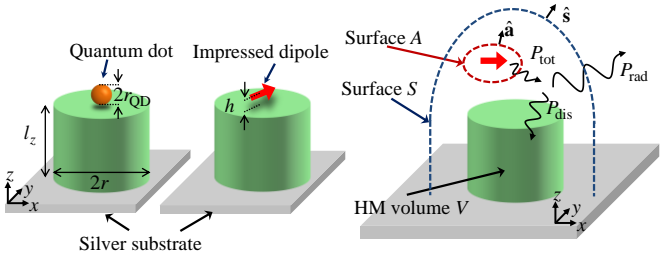

Fig. 1. Schematic view of a quantum dot (QD) emitter located on the top surface of a subwavelength HM resonator (left). The QD is modeled as an impressed dipole with an arbitrary polarization (middle). The right panel shows the representation of the power radiated into the upper free-space (through the surface S) which is the difference of the total emitted power by the dipole $P_{tot}$ (through the surface $A$) and the power dissipated in the HM resonator and substrate.

The QD modelled as a point dipole generates a very wide wavenumber-spectrum of waves enabling the emitted radiation to couple to modes in the HM resonator, and thus polarizes the HM resonator. Some excited mode is suitable for radiation enabling the use of the HM resonator as a nano-antenna. Part of the total emitted power $P_{tot}$ is dissipated as heat due to material losses (non-radiative process) in the HM, whereas the rest propagates away as radiation into the upper free space, $P_{rad}$. In this paper, we focus on enhancing the radiated power into the upper free-space, hence it is useful to define the radiative emission enhancement (*REE*) factor for each frequency as

$$REE = P_{rad} / P_{fs} \qquad (1)$$

where $P_{fs}$ is the power emitted by an isolated dipole in free space (vacuum). The *radiated* power $P_{rad}$ is evaluated by integrating the real part of the Poynting vector over the surface $S$:

$$P_{rad} = \frac{1}{2} \int_S \text{Re}\left(\mathbf{E} \times \mathbf{H}^*\right) \cdot \hat{\mathbf{s}} dS . \qquad (2)$$

On the other hand, the total power *emitted* by the dipole, which includes the one dissipated in the materials and the radiated one into the upper space, is similarly evaluated as

$$P_{tot} = \frac{1}{2} \oint_A \text{Re}\left(\mathbf{E} \times \mathbf{H}^*\right) \cdot \hat{\mathbf{a}} dA , \qquad (3)$$

where the surface $A$ is shown in Fig. 1. The dissipated power is $P_{dis} = P_{tot} - P_{rad}$, and can also be evaluated as

$$P_{dis} = \frac{1}{2} \int \text{Re}\left[\mathbf{E} \cdot \left(j\omega\mathbf{D}\right)^*\right] dV , \qquad (4)$$

where $V$ is the volume of the substrate and the resonator. Most of previous work involving HMs focused on boosting the emission rate, i.e., on enhancing $P_{tot}$. Here, the quantity of interest is $P_{rad}$, the portion of $P_{tot}$ that is radiated into the upper space. In the following, we evaluate these power quantities via full wave simulations based on the frequency-

domain finite-element method[19]

The HM nano-antenna is designed to resonate at the wavelength of 660 nm, which is accessible using CdSe QDs. The multilayer HM resonator comprises silver (Ag) and silica glass (SiO$_2$) layers (Fig. 2). Measured relative permittivity functions of the constituent materials are used for modeling the materials and , at the wavelength of interest, $\lambda_0 = 660$ nm, one has $\varepsilon_{Ag} = -17.19 - j0.69$ and $\varepsilon_{SiO_2} = 2.12$. The resonator is designed as a cylinder made of three pairs of alternating layers of SiO$_2$ (the bottom layer) and Ag (each with a thickness of 12.5 nm) and the top is capped with 5-nm-thick SiO$_2$ to protect Ag. The resonators are deposited on top of a Ag-cladded substrate as in Fig. 2(a). The resonator has a radius of 54 nm and a total height of 80 nm. Based on previous studies, three pairs of metal-dielectric planar structure are sufficient to exhibit hyperbolic metamaterial properties[2].

The calculated radiated power into the upper free space $P_{rad}$, as well as $P_{dis}$ and $P_{tot}$ versus wavelength for an *x*-directed dipole are shown in Fig. 2(b). The dipole is located at a height $h = 10$ nm above the top center of the resonator. These powers are normalized to $P_{fs}$. $P_{tot}$ is enhanced by 30 folds due to Purcell effect, the radiative emission and dissipated power are enhanced by 19 (=*REE*) and 9 times, respectively, at the antenna resonance, peaking around 660 nm. We stress that a large portion of the total power emitted is actually radiated into upper free-space. It is convenient to define the radiation efficiency as the fraction of radiated power over the total emitted one, $P_{rad} / P_{tot}$ (this coincides with the analogous definition used in antenna theory). Results in Fig. 2(b) show that $P_{rad} / P_{tot}$ is approximately 66% indicating that HM subwavelength resonators can be used as efficient nano-antennas. Using an alternative metric, for example as that in [16], the free-space radiation enhancement can be defined as the radiated power normalized to that radiated by the same dipole either over a flat Ag or over a HM substrate, leading to even larger enhancement factors of 102 and 143, respectively (plots not shown here for brevity).

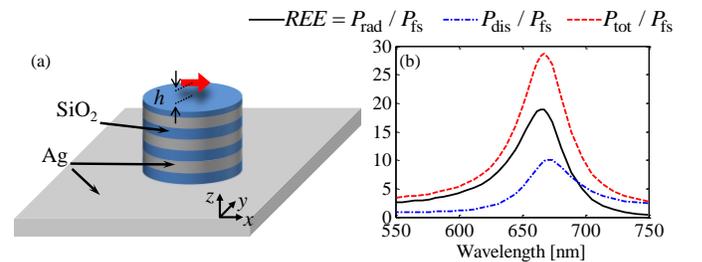

Fig. 2. (a) The illustration of the multilayer HM resonator on top of silver substrate. (b) Enhancement of radiated, dissipated and total power emitted relative to the power of the same dipole radiated in free-space ( $P_{fs}$ ).

Now we explore how the dipole orientation affects the *REE*. In Fig. 3, *REE* is plotted for dipoles polarized along *x*, *y*, and *z*, showing maximum *REE* of 25, 20 and 100, respectively. In addition, we also find that *REE* is sensitive to the location of the dipole relative to the HM resonator.

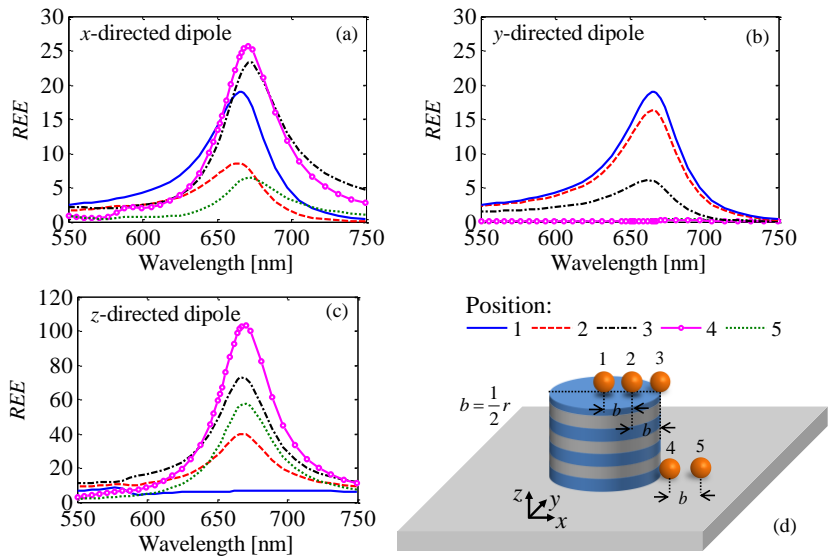

Fig. 3. Radiative emission enhancement *REE* of a QD versus wavelength model as a single dipole emitting with (a) *x*-, (b) *y*-, (c) *z*-polarization, for the enumerated QD positions in (d) (position 1: blue solid line, 2: red dashed line, 3: black dotted-dashed line, 4:, purple line with circle markers and 5: green dotted line)

The dipole positions explored are denoted by numbers $N = \{1, 2, 3, 4, 5\}$ in Fig. 3 (d). For the *x*-directed dipoles at positions 1, 3, and 4 as well as for *y*-directed dipoles at positions 1 and 2 the enhancement factors are larger than 15 folds. On the other hand, for a *z*-directed dipole, all positions except position 1 yield enhancement factors larger than 30. The maximum enhancement is achieved for the *z*-directed dipole at position 4 with an enhancement factor larger than 100. Using the effective medium approach[1,11] (EMA) to model the HM nano-antenna as a bulk anisotropic material, simulations are found to be 5% larger for the resonance wavelength and %20 smaller for the *REE*. In Fig. 4, we report the far-field directivity patterns pertaining to dipoles polarized along *x* and *z* at positions 1 and 4, respectively. Here directivity is defined as $\text{Directivity}(\theta, \phi) = U(\theta, \phi) 4\pi / P_{\text{rad}}$ where *U* is the radiation intensity per unit solid angle. The HM nano-antenna radiates with a single lobe in the +ve *z* direction and it is nearly symmetric about the *z*-axis with a half-power beam width of 95°. The maximum directivity is 5.7, which is almost 4 times that of a dipole in vacuum (For a point dipole, the maximum directivity is 1.5). Therefore the field intensity in +z direction is further enhanced due to enhanced directivity. In addition, the radiation patterns obtained using EMA are found to be in very good agreement with those using the multilayer HM simulations.

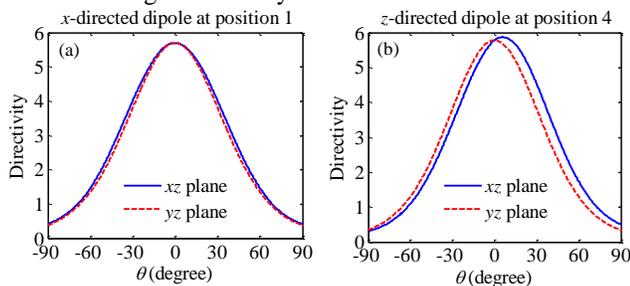

Fig. 4. Directivity patterns of (a) *x*-directed dipole at position 1 and (b) *z*-directed dipole at position 4.

In order to provide a physical insight into this characteristic,

in Fig. 5 we report the instantaneous electric field vector map at the 660 nm resonance when the resonator is excited by an *x*-directed dipole at position 1; the arrows indicate the E-field direction and the color indicates its magnitude. A similar E-field distribution is found for the *z*-directed dipole at position 4, with the only difference being that the E-field is much stronger. The vector field map shown in Fig. 5 is basically the modal field of the resonance. From these field maps, we can correlate the emission enhancement factor to the position and dipole orientation. For example, when a *z*-directed dipole is place in location 4 where the *z*-component of the field is strong, the maximum enhancement factor is obtained. In contrast, the enhancement factor of the *x*-directed dipole at position 1 is smaller than that of the *z*-directed dipole at position 4, due to the fact that the *x*-component of the modal E-field in this position is weaker. Instead, the *x*-oriented dipole at location 3 matches the polarization of the modal E-field and therefore it exhibits a comparable enhancement factor as the *z*-oriented dipole at location 4. These results correlate with an increase of the local density of states. One can envision that further enhancement may be realized using different geometrical shapes that produce even stronger modal E-field intensities where a QD can be positioned.

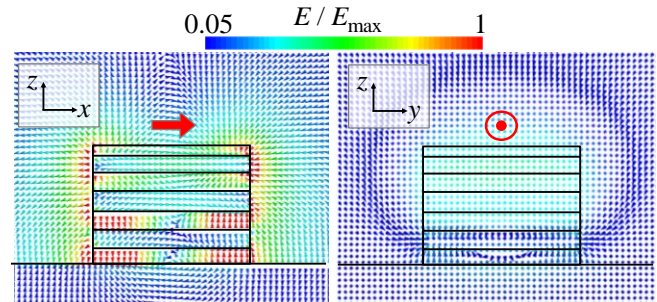

Fig. 5. Normalized electric field vector maps in the two principal planes at 660 nm wavelength. The color bar is in linear scale. The dipole is polarized along the *x*-axis as illustrated. Note that the field vector lies mainly on the *xz* plane.

In a practical implementation, HM resonators are fabricated in a two-dimensional array and the QDs will be dispersed on the surface hosting the array. Here we present a discussion on how nearby HM resonators affect the radiative emission enhancement. As mentioned earlier, an advantage of using HM resonators lies in the opportunity of high packing density due to their small size. In full-wave simulations the implementation of an infinitely periodic array with a single dipole excitation is impractical from numerical simulation standpoint even when taking advantage of the structure periodicity [20], therefore in the following we analyze a finite array of resonators. In particular we investigate a 3-by-3 square array of resonators identical to the one previously introduced and show the radiation enhancement pertaining to an $x$-directed dipole at position 1 and a $z$-directed dipole at position 4, as depicted in Fig. 6. The enhancement of radiated power *REE* is plot in Fig. 7 for four different periods $d_x = d_y$: 175, 200, 225, and 250 (all in nm).

As a comparison we also provide the enhancement for an isolated HM resonator (no-array).

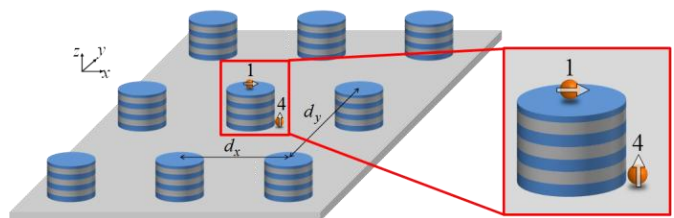

Fig. 6. A 3-by-3 array of cylindrical subwavelength HM resonators with a QD located in two possible positions.

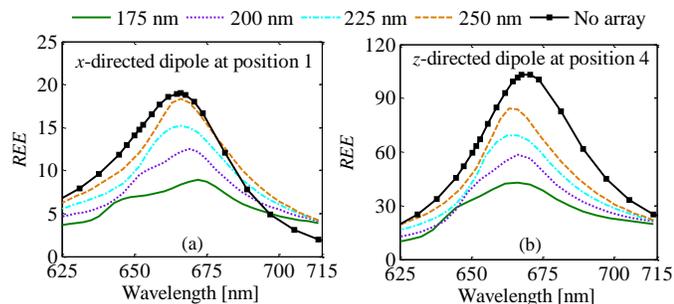

Fig. 7. Radiative emission enhancement *REE* versus wavelength for (a) an $x$-directed dipole at position 1 and (b) for a $z$-directed dipole at position 4, periods of $d_x = d_y = 175$ nm (green solid line), 200 nm (purple dotted line), 225 nm (cyan dotted-dashed line), and 250 (orange dashed line) as in Fig. 6, and for the "no-array" case (black solid line with square markers).

Within the considered parameter range, the enhancement increases as the period $d_x = d_y$ is increased as shown in Fig. 7. For $d_x = 200$ nm ($\lambda_0 / 3.3$) the enhancement values are still high, and more than 50% of that of the no-array case. The case with largest spacing $d_x = 250$ nm ($\lambda_0 / 2.6$, i.e., still subwavelength) yields a level of enhancement comparable to the no-array case. This implies that HM resonators can be densely arranged in array configuration with subwavelength period still keeping large power enhancement levels. As in the single resonator case, the $z$-oriented dipole at position 4 in the array yields much higher *REE* than the other locations. For randomly oriented QDs, therefore, the average enhancement effect is expected to be smaller.

We have shown for the first time, to the authors' knowledge, that subwavelength HM resonators can be used as efficient radiators (i.e., as efficient optical nano-antennas). They have the ability to enhance both the radiative emission and the Purcell factor of quantum emitters in their vicinity. We have shown that the total emission of a QD near a HM resonator nano-antenna is mostly coupled to the radiated power into the upper free space. The results show up to 100 times free-space radiation enhancement in the presence of HM resonators, indicating that these HM resonators are efficient radiators. The enhancement strongly depends on the dipole position and polarization. The combined effect of the enhanced REE and the directive emission produces further enhancement of field intensity in the broad side up to 400 times. Moreover significant enhancement levels are achieved also in the presence of surrounding nano-antennas in a two-dimensional array configuration. We have shown that the presence of nearby HM resonators in a 3-by-3 square array lead to large power enhancement levels, especially for QD at position 4 (i.e. for QD sitting next to the resonator, on top of the substrate), slightly lower than those for a single HM resonator. We conclude that nanostructures made of HM may prove useful for improvement of light-matter interaction and solid-state lighting applications.

This work was performed, in part, at the Center for Integrated Nanotechnologies, a U.S. Department of Energy, Office of Basic Energy Sciences user facility. The computational result was supported by Sandia LDRD program and the permittivity measurements performed at Sandia were supported by Sandia's Solid-State Lighting Science EFRC. Sandia National Laboratories is a multi-program laboratory managed and operated by Sandia Corporation, a wholly owned subsidiary of Lockheed Martin Corporation, for the U.S. Department of Energy's National Nuclear Security Administration under contract DE-AC04-94AL85000. F.C. and C.G. are grateful to Ansys Inc. for providing HFSS.